\documentclass[12pt]{iopart}
\usepackage{iopams}
\usepackage{graphicx}
\begin{document}

\title[Garza-Soto et al. 2020]{Geometric-phase polarimetry}

\author{Luis Garza-Soto, Alejandra De-Luna-Pamanes,  Israel Melendez-Montoya, Natalia Sanchez-Soria, Diana Gonzalez-Hernandez, and Dorilian Lopez-Mago$^\ast$}

\address{Tecnologico de Monterrey, Escuela de Ingenier\'{i}a y Ciencias, Ave. Eugenio Garza Sada 2501, Monterrey, N.L. 64849, Mexico.}
\ead{dlopezmago@tec.mx}
\vspace{10pt}
\begin{indented}
\item[]August 2020
\end{indented}

\begin{abstract}
This paper describes polarimetric strategies based on measuring the light's geometric phase, which results from the evolution of the polarisation state while traversing an optical system. The system in question is described by a homogeneous Jones matrix, which by definition, contains mutually perpendicular eigenpolarisations. Our leading theory links the system's Jones matrix parameters (eigenvalues and eigenvectors) with the input polarisation state and the geometric phase. We demonstrate two interferometric techniques. The first one measures the geometric phase based on the relative lateral fringe displacement between the interference pattern of two mutually-orthogonal polarisation states. The second technique uses the visibility of the interference fringes to determine the eigenpolarisations of the system. We present proof-of-principle experiments for both techniques. 
\end{abstract}

%
%
%
%
%

\section{Introduction}

A point on the Poincar\'{e} sphere represents the polarisation state of a monochromatic light beam. If the light beam propagates through optical elements that change its polarisation state, the beam acquires not only a dynamic phase from the optical path length but also a geometric or Pancharatnam-Berry phase. This geometric phase depends on the trajectory connecting the initial and final points on the Poincar\'{e} sphere \cite{Berry1984,Berry1987}.

The Pancharatnam-Berry phase has applications in the area of structured light, which studies the generation of custom optical fields and their interaction with structured materials, such as Pancharatnam-Berry phase optical elements (PBOEs). These PBOEs can be used for wavefront shaping \cite{Marrucci2006}, manipulation of vector beams \cite{Zhou2016,Wang2018} and waveguides that operate entirely on geometric phases \cite{Slussarenko2015}. Interferometric techniques can be used to measure the geometric phase, as long as the dynamical phase is properly separated~\cite{Taylor1992,Loredo2009,Kobayashi2010,VanDijk2010a,Lopez-Mago2017a}. The geometric phase has been experimentally demonstrated for classical light beams and even for quantum states of light \cite{Kwiat1991}.

Optical polarimetry encompasses the theory and techniques to measure light's polarisation state and its interaction with polarising optical systems. Jones calculus studies the interaction of polarised light with polarisation systems. The Jones matrix of a polarising optical system provides information about its polarisation properties \cite{Savenkov2007}. Its eigenvectors or eigenpolarisations are those polarisation states of light that do not change after passing the polarisation system. More precisely, the input and output polarisation states are equal, but the amplitude and overall phase of the output electric field changes. The corresponding eigenvalues or complex transmittances characterise these changes. Jones matrices are classified according to the orthogonality of their eigenpolarisations. If the eigenpolarisations are orthogonal, the system is homogeneous; otherwise, the system is inhomogeneous \cite{Lu1994}.

Most common polarimetric methods rely on intensity projections \cite{GoldsteinBook, collett2005field,ChipmanBook}. On the contrary, in this work, we present two polarimetry techniques that rely on the geometric phase. Our techniques are interferometric and can be used to determine the eigenpolarisations and eigenvalues of an optical system's Jones matrix. The two methods share the same basic theory, which is based in our previous work \cite{Lopez-Mago2017a}, and further supported by the recent theoretical work of Gutierrez-Vega \cite{Gutierrez-Vega2020-1,Gutierrez-Vega2020-2}. We consider, in all techniques, polarising optical elements characterised by homogeneous Jones matrices.

This article is organised as follows. Section \ref{Sec:theory} establishes the main theory, formulae and notations. Section \ref{Sec:fringe_shifting} presents the first method, which consists of measuring the geometric phase using a fringe-shifting interferometer. We use this method to measure the retardance of an exemplary system: a variable linear retarder composed of three waveplates. Also, we derive an alternative formulation, based entirely on the geometric phase, to calculate the Stokes parameters of an unknown polarisation state. Section \ref{Sec:fringe_contrast} shows the second method, which performs visibility measurements to infer the geometric phase. We provide a functional relationship between the visibility or contrast of the interference fringes with the geometric phase introduced by the system. Discussion and conclusions are drawn in section \ref{Sec:conclusions}.

\section{The geometric phase of homogeneous Jones matrices} \label{Sec:theory}

We consider a polarising optical system characterised by a homogeneous Jones matrix $\mathbf{J}$. The matrix has two orthonormal eigenpolarisations described by the $2\times 1$  Jones vectors
\begin{eqnarray}
    \mathbf{E}_{1} = \left( \matrix{ q_{x} \cr q_{y}} \right), \quad  \mathbf{E}_{2}= \left( \matrix{-q_{y}^{\ast} \cr q_{x}^{\ast}}\right),
\end{eqnarray}
where $q_{x},q_{y} \in \mathbb{C}$ and $|q_{x}|^{2}+|q_{y}|^{2}=1$. The corresponding normalised Stokes vectors are $\mathbf{Q}_{1}$ and $\mathbf{Q}_{2}$. The orthogonality condition means that $\mathbf{E}_{1}\cdot \mathbf{E}_{2}^{\ast}=0$ and $\mathbf{Q}_{1}=-\mathbf{Q}_{2}=\mathbf{Q}$, with $\mathbf{Q}=(Q_{1};Q_{2};Q_{3})$. The eigenvalues of $\mathbf{J}$ are $\mu_{1},\mu_{2}\in \mathbb{C}$, therefore $\mathbf{J} \mathbf{E}_{1}=\mu_{1} \mathbf{E}_{1}$ and $\mathbf{J} \mathbf{E}_{2}=\mu_{2} \mathbf{E}_{2}$. If light with polarisation $\mathbf{E}_{a}$ passes through $\mathbf{J}$, the state of the resulting field is given by $\mathbf{E}_{b}=\mathbf{J}\mathbf{E}_{a}$. According to the Pancharatnam connection \cite{Berry1987}, the phase difference $\Phi$ between the polarization states $\mathbf{E}_a$ and $\mathbf{E}_b$, is equal to 
\begin{equation}
    \Phi = \mathrm{arg} \left[ \mathbf{E}_{a}^{\ast} \cdot \mathbf{E}_b \right] = \mathrm{arg} \left[ \mathbf{E}_{a}^{\ast} \cdot \mathbf{J}\mathbf{E}_a \right], \label{Eq:totalphase1}
\end{equation}
where $(\cdot)$ stands for the dot product of vectors. We have demonstrated that the total phase difference between $\mathbf{E}_{a}$ and $\mathbf{E}_{b}$ can be written as \cite{Lopez-Mago2017a,Gutierrez-Vega2011a,Melendez-Montoya2018}
\begin{equation} 
\Phi = \mathrm{arg} \left[\mu_{1} + \mu_{2} + (\mu_{1}-\mu_{2})\mathbf{Q}\cdot \mathbf{A} \right], \label{Eq:totalphase}
\end{equation}
where $\mathbf{A}$ is the normalised Stokes vector for $\mathbf{E}_{a}$. Equation (\ref{Eq:totalphase}) may be split into two phases, $\Phi=\Phi_{\mathrm{D}} + \Phi_{\mathrm{G}}$, where $\Phi_{\mathrm{D}}=\mathrm{arg}(\mu_{1}\mu_{2})/2$ is the dynamic phase, and $\Phi_{\mathrm{G}}$  is the geometric phase. 

From an inspection of equation (\ref{Eq:totalphase}), it becomes evident that we can extract the polarisation properties of $\mathbf{J}$ (i.e., $\mu_1$, $\mu_2$ and $\mathbf{Q}$) through phase measurements. In other words, by measuring $\Phi$ as a function of $\mathbf{A}$, we can find the eigenvectors $\mathbf{Q}_{1},\mathbf{Q}_{2}$ and eigenvalues $\mu_{1},\mu_{2}$ of the polarising optical system. With that information, the Jones matrix  can be written as~\cite{Gutierrez-Vega2011a}
\begin{equation}\label{Eq:JonesMatrix}
\mathbf{J} = \left( \matrix{
\mu_{1}|q_{x}|^{2} + \mu_{2}|q_{y}|^{2} & (\mu_{1}-\mu_{2})q_{x}q_{y}^{\ast} \cr (\mu_{1}-\mu_{2})q_{x}^{\ast}q_{y} & \mu_{2}|q_{x}|^{2} + \mu_{1}|q_{y}|^{2} }\right).
\end{equation}

An important observation of our previous work is that the geometric phase $\Phi_{\mathrm{G}}$ acquired by $\mathbf{A}$ is opposite in sign to the geometric phase $\Phi_{\mathrm{G}}^{\perp}$ acquired by its orthogonal state $\mathbf{A}^{\perp}=-\mathbf{A}$, but the dynamic phase is equal for both \cite{Lopez-Mago2017a}. It means that $\Phi^{\perp} = \Phi_{\mathrm{D}}-\Phi_{\mathrm{G}}$. Therefore, $\Phi-\Phi^{\perp} =2\Phi_{\mathrm{G}}$ and
\begin{eqnarray}\label{Eq:PBphase1}
\Phi_{\mathrm{G}} = \frac{1}{2} \left\{ \mathrm{arg} \left[ \mu_{1} + \mu_{2} + (\mu_{1}-\mu_{2})\mathbf{Q}\cdot \mathbf{A} \right] - \right. \nonumber \\
 \left. \mathrm{arg} \left[ \mu_{1} + \mu_{2} - (\mu_{1}-\mu_{2})\mathbf{Q}\cdot \mathbf{A} \right] \right\} . 
\end{eqnarray}

In what follows, we neglect the thickness of the polarisation system, which means that the arguments of $\mu_{1}$ and $\mu_{2}$  are purely geometric phases. Of course, in the experimental implementations, we separate the dynamic phase from the geometric phase. Furthermore, for homogeneous and non-absorbing polarisation systems $|\mu_{1}|=|\mu_{2}|=1$ \cite{Savenkov2007}. Thus, we write $\mu_{1}=\exp(i \delta)$ and $\mu_{2}=\exp(-i\delta)$. Now, equation (\ref{Eq:PBphase1}) can be reduced to
\begin{equation} 
\tan (\Phi_{\mathrm{G}}) =  \mathbf{Q}\cdot \mathbf{A} \tan(\delta). \label{Eq:tanphiPB}
\end{equation}

The above equation is the first significant result of this work. It gives a simple expression for the geometric phase introduced by homogeneous Jones matrices. Besides, it provides a method to measure either the eigenpolarisations of the optical system or the normalised Stokes parameters of an unknown polarisation state.

To determine the eigenvalues and eigenvectors, we notice that equation (\ref{Eq:tanphiPB}) has three unknowns: $\delta$ and two components of $\mathbf{Q}$, given that $Q_{1}^{2}+Q_{2}^{2}+Q_{3}^{2}=1$. Therefore, in theory, we need at least three measurements of $\Phi_{\mathrm{G}}$ in order to determine the eigenvalues and eigenvectors. 

We do the following procedure to determine $\mathbf{Q}$ and $\delta$ from equation (\ref{Eq:tanphiPB}). We perform three measurements with three different polarisation states given by the Stokes vectors $\mathbf{A}, \mathbf{B},\mathbf{C}$. Then, we define $\mathbf{\bar{Q}}=\tan(\delta) \mathbf{Q}$, and solve the resulting system of equations
\begin{equation}\label{Eq:matrixtogetQ}
\left( \matrix{
A_{1} & A_{2} & A_{3} \cr B_{1} & B_{2} & B_{3} \cr C_{1} & C_{2} & C_{3}
}\right) \left( \matrix{
\bar{Q}_{1} \cr \bar{Q}_{2} \cr \bar{Q}_{3} 
}\right) = \left( \matrix{
\tan (\Phi_{\mathrm{G}})_{1} \cr \tan (\Phi_{\mathrm{G}})_{2} \cr \tan (\Phi_{\mathrm{G}})_{3}
}\right),
\end{equation}
in order to determine $\mathbf{\bar{Q}}$. Finally, we obtain $\delta$ from $\tan \delta = |\mathbf{\bar{Q}}|=\sqrt{\bar{Q}_{1}^{2}+\bar{Q}_{2}^{2}+\bar{Q}_{3}^{2}}$.

The polarisation states that we use to solve equation (\ref{Eq:matrixtogetQ}) are the basis polarisation states $\mathbf{A} = (1;0;0)$, $\mathbf{B}=(0;1;0)$ and $\mathbf{C}=(0;0;1)$, which are the horizontal, diagonal and circular (right handed) polarisation states, respectively.

\section{Fringe-shifting polarimetry}\label{Sec:fringe_shifting}

\begin{figure}[ht!]
\centering\includegraphics[width=8cm]{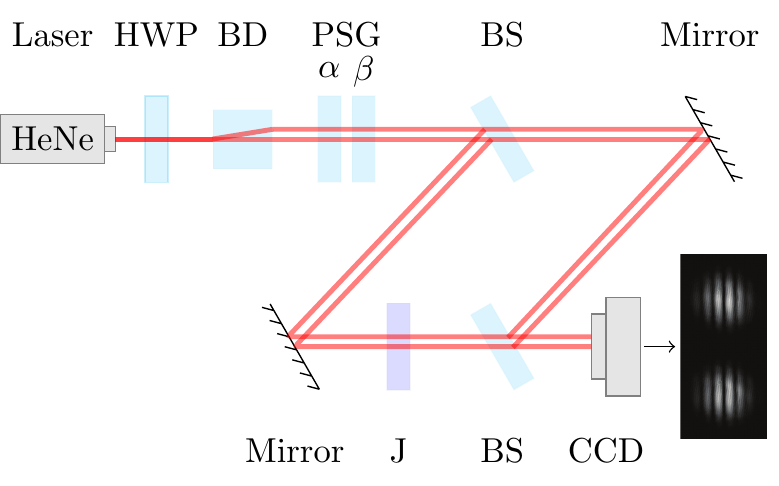}
\caption{Schematic of our experimental arrangement to measure the geometric phase through fringe-shifting interferometry. Laser source: He-Ne laser. \textbf{BD}: beam displacer. \textbf{PSG}: polarisation state generator, which is composed of a quarter-wave plate and a half-wave plate with their fast axes oriented an angle $\alpha$ and $\beta$, respectively.  \textbf{BS}: plate beam splitter. \textbf{CCD}: camera. $\mathbf{J}$: polarising optical system. The shape of the interferometer is used to minimise the polarisation changes from an oblique incidence.}\label{Fig:experiment}
\end{figure}

Our first polarimetric method uses fringe-shifting interferometry to measure the geometric phase from the relative fringe displacement between two orthogonal input states of polarisation. Figure \ref{Fig:experiment} shows a schematic of our experimental arrangement. The optical setup is composed of three parts: (i) the polarisation state generator (PSG), (ii) a Mach-Zehnder (MZ) interferometer of rhomboidal shape, and (iii) the measurement-analysis system.

\subsection{The polarisation state generator}

The PSG consists of a linearly polarised He-Ne laser, a half-wave plate and a beam displacer followed by a pair of retarders: a quarter-wave plate $\mathbf{QWP}(\alpha)$ and another half-wave plate $\mathbf{HWP}(\beta)$. The beam displacer generates two parallel beams with orthogonal polarizations. One beam is the ordinary polarized ($\mathbf{o}$), and the other is the extraordinary polarized ($\mathbf{e}$). The beam displacer is oriented such that the $\mathbf{o}$ and $\mathbf{e}$ modes correspond to the vertical and horizontal polarisation states, respectively. The half-wave plate previous to the beam displacer is used to fine-tune the splitting, such that both beams have equal power.  Our beam displacer is a calcite prism (Thorlabs BD40) that provides a beam separation of $4$ mm. We also implemented a tunable beam displacer \cite{Salazar-Serrano2015}, but found some stability issues and decided to keep the calcite prism. The beam separation should be enough to avoid the overlap between the beams. Since our beam diameter is about $1$ mm, the beam displacing prism works fine.  

The plates $\mathbf{QWP}(\alpha)$ and $\mathbf{HWP}(\beta)$ are mounted in a fully-automated rotation mount. By properly orienting their fast axes, we transform the input beams into two mutually-orthogonal elliptical polarisation states. By using the Jones matrices \cite{collett2005field}
\begin{equation} \label{Eq:Qmatrix}
\mathbf{QWP}(\alpha) = \frac{1}{\sqrt{2}} \left( \matrix{ 1+i\cos(2\alpha) & i\sin(2\alpha) \cr i\sin(2\alpha) & 1-i\cos(2\alpha)          }\right),
\end{equation}
\begin{equation}\label{Hmatrix}
\mathbf{HWP}(\beta) =i\left( \matrix{
\cos(2\beta) & \sin(2\beta) \cr \sin(2\beta) & -\cos(2\beta)
}\right),
\end{equation}
we can show that light with horizontal polarisation passing through $\mathbf{HWP}(\beta)\mathbf{QWP}(\alpha)$ will come out with a polarisation state described by the normalised $3 \times 1$ Stokes vector
\begin{equation}\label{Eq:Stokesafterbeamgenerator}
\mathbf{S}^{\mathrm{h}} = \left( \matrix{  \cos(2\alpha)\cos(4\beta-2\alpha) \cr \cos(2\alpha)\sin(4\beta-2\alpha)\cr - \sin(2\alpha) }\right).
\end{equation}
Similarly, we can show that light with vertical polarisation will exit as $\mathbf{S}^{\mathrm{v}}=- \mathbf{S}^{\mathrm{h}}$. It can be seen from equation (\ref{Eq:Stokesafterbeamgenerator}) that $\mathbf{S}^{\mathrm{h}}$ has the form of a position vector in spherical coordinates and thus it spans all points on the surface of the Poincar\'{e} sphere.  

\subsection{The rhomboidal-shape Mach-Zehnder interferometer}

Both beams are introduced in the MZ interferometer in such a way that each one interferes with itself. Two copies of each state are created when passing through the 50:50 non-polarising plate beam splitter (BS). One of the arms of the interferometer contains the polarisation system $\mathbf{J}$ under study. At the second BS, we have the interference between $\mathbf{E}_{a}$ and $\mathbf{E}_{b}=\mathbf{J}\mathbf{E}_{a}$ for each of the input polarisation states. 

The rhomboidal shape of the interferometer is used to reduce the unwanted polarisation changes coming from non-normal incidence on the BS \cite{Ericsson2005}. We attribute the polarisation changes to phase differences between the vertical (TE) and horizontal (TM) modes at the BS after reflection or transmission \cite{Pezzaniti:94}. In order to determine the incidence angle, we monitor the reflection of a diagonally-polarised beam using a commercial polarimeter (Thorlabs PAX1000VIS). We choose the incidence angle in which the polarisation of the reflected beam equals the input beam.

We align the interferometer to obtain transverse interference fringes by adding a tilt to one of the mirrors. The observed interference pattern, for one of the input beams, can be described in terms of the axis perpendicular to the fringes, let's say $x$, as \cite{Lopez-Mago2017a}
\begin{equation}
   I(x) \propto 1 + V \cos(\kappa x+\Phi_{\mathrm{D}}+\Phi_{\mathrm{G}}), \label{Eq:interfpattern1}
\end{equation}
where $V$ and $\kappa$ are the visibility and spatial frequency. Correspondingly, the interference pattern of the accompanying beam is
\begin{equation}
   I^{\perp}(x) \propto 1 + V \cos(\kappa x+\Phi_{\mathrm{D}}-\Phi_{\mathrm{G}}). \label{Eq:interfpattern2}
\end{equation}
By comparing equations (\ref{Eq:interfpattern1}) and (\ref{Eq:interfpattern2}), it becomes clear that the relative fringe displacement contains the geometric phase $\Phi_{\mathrm{G}}$.

\subsection{The measurement system}

We record the interference pattern using a CCD camera (Thorlabs DCU223). We process the raw data to eliminate background noise introduced by the optical elements and the camera. Since the noise features have high frequencies compared to the frequency of the fringes, we eliminate them using a digital low-pass spatial frequency filter on the image of the interference pattern. We accomplish this by applying a Fourier transform to the image and multiplying it by a spatial filter (specifically, a two-dimensional Gaussian filter). We then apply an inverse Fourier transform to recover the filtered image \cite{Gonzalez:2003:DIP:993475}. 

For each interference pattern (see the inset of figure \ref{Fig:experiment}), we take the mean profile of the fringes by averaging a region located at the centre of each interferogram. As a result, we get the interferograms equivalent to $I(x)$ and $I^{\perp}(x)$ in equations (\ref{Eq:interfpattern1}) and (\ref{Eq:interfpattern2}). The relative shift between $I(x)$ and $I^{\perp}(x)$ is obtained by applying the following Fourier analysis.

\subsection{Numerical analysis}

We use the following numerical analysis to extract the geometric phase from our measurements. The Fourier transform of $f(x)$ is defined by 
\begin{equation}
\hat{f}(k) = \frac{1}{\sqrt{2\pi}} \int _{-\infty}^{\infty} f(x) e^{-i k x} \mathrm{d} x.
\end{equation}
It is well-known that the Fourier transform of $f(x)\cos(k_{0}x)$ is
\begin{equation}\label{Eq:Fourieridentitycosine}
\hat{f}(k) = \frac{1}{2} \left[ \hat{f}(k-k_{0})+\hat{f}(k+k_{0})\right].
\end{equation}

In our measurements, we have an interference pattern of the form $F(x) \approx G(x)\cos(\Phi + k_{0}x)$, where $G(x)$ is a Gaussian envelope. By defining $\Phi = \Phi_{D} + \Phi_{\mathrm{G}}$, we can apply the sum identity for the cosine function to get 
\begin{equation}
F(x) \approx G(x)[ \cos(k_{0}x)\cos(\Phi) - \sin(k_{0}x)\sin(\Phi)]. 
\end{equation}
Applying equation (\ref{Eq:Fourieridentitycosine}) to the previous expression we obtain
\begin{eqnarray}
\hat{f} (k) \approx & \frac{1}{2} \cos\Phi \left[ \hat{f}(k - k_{0}) + \hat{f}(k + k_{0})  \right] - \nonumber \\ & \sin \Phi \frac{1}{2i} \left[ \hat{f}(k - k_{0})-\hat{f}(k + k_{0}) \right].
\end{eqnarray}
Arranging common terms,
\begin{equation}
\hat{f}(k) \approx \hat{f}(k - k_{0})\frac{1}{2}\left[ \cos\Phi + i \sin\Phi \right] + \hat{f}(k + k_{0})\frac{1}{2}\left[ \cos\Phi - i\sin\Phi \right],  
\end{equation}
which simplifies to
\begin{equation}
\hat{f}(k) = \frac{1}{2} \left[\exp(i\Phi) \hat{f}(k-k_{0}) + \exp(-i\Phi)\hat{f}(k+k_{0}) \right].
\end{equation}

In our case, we are working with two fields of the form
\begin{eqnarray}
F_{1}(x) &=& G(x)\cos(k_{0}x+\Phi_{D}+\Phi_{\mathrm{G}}), \\
F_{2}(x) &=& G(x)\cos(k_{0}x+\Phi_{D}-\Phi_{\mathrm{G}}).
\end{eqnarray}
The Fourier transform of each field results in 
\begin{eqnarray}
\hat{f}_{1}(k) = & \frac{1}{2} \left[ \exp[i(\Phi_{D}+\Phi_{\mathrm{G}})]\hat{f}(k-k_{0}) + \right. \nonumber \\ 
& \left. \exp[-i(\Phi_{D}+\Phi_{\mathrm{G}})] \hat{f}(k+k_{0})\right], \\
\hat{f}_{2}(k) =& \frac{1}{2} \left[ \exp[i(\Phi_{D}-\Phi_{\mathrm{G}})]\hat{f}(k-k_{0}) + \right. \nonumber \\ 
& \left. \exp[-i(\Phi_{D}-\Phi_{\mathrm{G}})] \hat{f}(k+k_{0})\right]. 
\end{eqnarray}
We multiply both functions to separate $\Phi_{\mathrm{G}}$ from $\Phi_{D}$. The result is 
\begin{eqnarray}
4\hat{f}_{1}^{\ast}\hat{f}_{2}  = & \exp(-i2\Phi_{\mathrm{G}}) \hat{f}^{\ast}(k - k_{0})\hat{f}(k - k_{0}) + \nonumber \\ 
& \exp(i2\Phi_{\mathrm{G}}) \hat{f}^{\ast}(k + k_{0})\hat{f}(k + k_{0}) + \nonumber \\ 
& \exp(-i2\Phi_{D}) \hat{f}^{\ast}(k - k_{0})\hat{f}(k + k_{0}) + \nonumber \\
& \exp(i2\Phi_{D}) \hat{f}^{\ast}(k + k_{0})\hat{f}(k - k_{0}).
\end{eqnarray}

Notice that $\hat{f}^{\ast}(k + k_{0})\hat{f}(k - k_{0})=0$ (the separation between $\hat{f}(k + k_{0})$ and $\hat{f}(k - k_{0})$ is larger than the width of their envelope). Hence, we can rewrite the expression as
\begin{equation}
\hat{f}_{1}^{\ast}(k) \hat{f}_{2}(k) = \frac{1}{4} \left[ \exp(-i2\Phi_{\mathrm{G}})|\hat{f}(k - k_{0})|^{2} + \exp(i2\Phi_{\mathrm{G}})|\hat{f}(k + k_{0})|^{2} \right].
\end{equation}

Finally, we evaluate the previous equation at $k=k_{0}$ and determine the phase of the resulting value, which is equal to $2\Phi_{\mathrm{G}}$. In practice, the function $|\hat{f}_{1}^{\ast}(k) \hat{f}_{2}(k)|$ contains two peaks located at $k=\pm k_{0}$. We localise those maxima to determine the value of $k_{0}$.

\subsection{Experimental results}

To corroborate our theory, we analyse the system $\mathbf{QWP}(\varphi_{1})\mathbf{HWP}(\varphi_{2})\mathbf{QWP}(\varphi_{1})$. This system is equivalent to a linear retarder where the orientation of the half-wave plate controls the amount of retardance between the eigenpolarizations. The retardance $R$ of a homogeneous polarisation system is defined as \cite{Lu1994}
\begin{equation}
R = | \arg(\mu_{1})-\arg(\mu_{2})|, \qquad 0\leq R \leq \pi. \label{Eq:retardance}
\end{equation}
Using matrices (\ref{Eq:Qmatrix}) and (\ref{Hmatrix}), we find the following eigenpolarisations:
\begin{equation}
\mathbf{E}_{1} = \frac{1}{\sqrt{2}} \left( \matrix{
(1-\sin 2\varphi_{1})^{1/2} \cr (1-\sin 2\varphi_{1})^{-1/2}\, \cos 2\varphi_{1} 
}\right),
\end{equation}
\begin{equation}
\mathbf{E}_{2} = \frac{1}{\sqrt{2}} \left( \matrix{
-(1-\sin 2\varphi_{1})^{-1/2} \, \cos 2\varphi_{1} \cr (1-\sin 2\varphi_{1})^{1/2}
}\right),
\end{equation}
with complex eigenvalues given by
\begin{equation}
\mu_{1} = -\exp(i2\left[\varphi_{1}-\varphi_{2}\right]) \, , \;\; \mu_{2} = \mu_{1}^{\ast}. \label{Eq:eigenvaluesQHQ}
\end{equation}
According to equation (\ref{Eq:retardance}), the retardance of this system is
\begin{equation}\label{Eq:RforJ}
R=|4\varphi_{1}-4\varphi_{2}|.
\end{equation}
Without loss of generality, we fix the value of $\varphi_{1} = 0$ and measure the retardance of this system as a function of $\varphi_{2}$. 

\begin{figure}[ht!]
\centering\includegraphics[width=10cm]{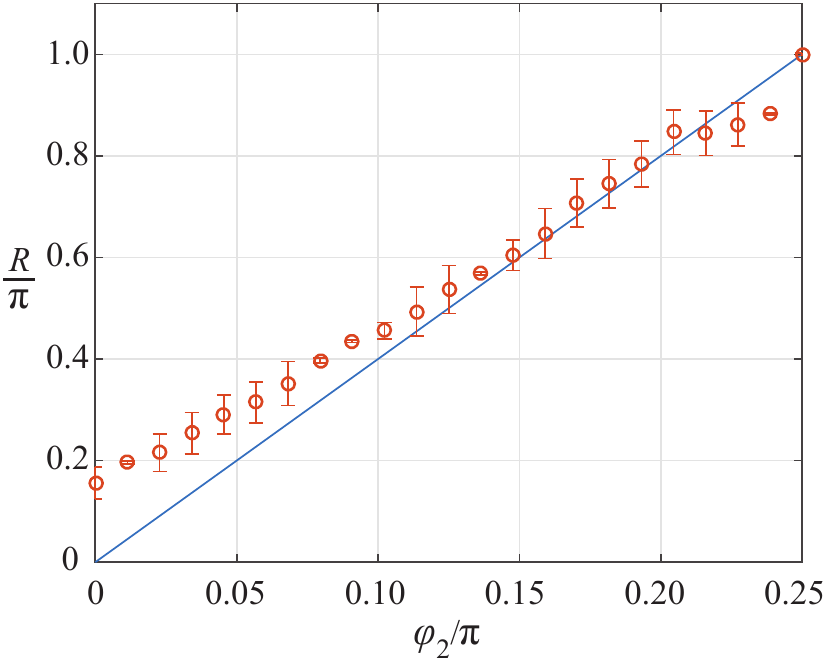}
\caption{Retardance of the homogeneous system $\mathbf{QWP}(\varphi_{1})\mathbf{HWP}(\varphi_{2})\mathbf{QWP}(\varphi_{1})$ as a function of $\varphi_{2}$. The solid line is the theoretical result from equation (\ref{Eq:RforJ}) with $\varphi_{1}=0$.}\label{Fig:plotretardance}
\end{figure}

Figure \ref{Fig:plotretardance} shows our experimental results. We use equation (\ref{Eq:matrixtogetQ}) to calculate the system's eigenvalues. Then, we calculate the retardance using equation (\ref{Eq:retardance}). Our input polarisation states are the basis polarisations, i.e., horizontal, diagonal and circular. As seen in figure \ref{Fig:plotretardance}, our experimental results do not exactly match the theoretical calculations. We attribute these imperfections to different reasons: (1) the polarisation state generator does not produce the exact polarisation state due to inaccurate positioning of our rotating mounts, (2) the beams pass through different points on the retarder (close to the border, due to its spatial separation introduced by the beam displacer), alignment of the centroids of the beams, and orientation of the fringe pattern. In addition, when rotating the plates, the beams acquire a slight change on propagation directions, which changes the location of the centroids and introduces a phase error. Notwithstanding the lack of perfect agreement, we believe our results show the desired objective: we can use the geometric phase to measure polarimetric parameters. 

\subsection* {Geometric-phase-based Stokes parameters}

This subsection explains a unique formulation to determine the Stokes parameters using the geometric phase. We found interesting correspondence between this formulation and the common intensity-based Stokes measurements, which motivates us to include this analysis. However, this subsection can be left aside without affecting the understanding of our methods.

Contrary to intensity-based Stokes parameters, equation (\ref{Eq:tanphiPB}) reveals that we can obtain the Stokes parameters from geometric-phase measurements. In other words, we can determine the polarisation state of $\mathbf{A}$ by controlling the values of $\mathbf{Q}$. 

We consider three different polarisation systems with eigenpolarisations represented by the $3 \times 1$ Stokes vectors $\mathbf{P}$, $\mathbf{Q}$, and $\mathbf{R}$, with eigenvalues $\delta_{P}$, $\delta_{Q}$, and $\delta_{R}$, respectively. Applying equation (\ref{Eq:tanphiPB}) and defining $\mathbf{\bar{P}}=\mathbf{P}\tan\delta_{P}$, $\mathbf{\bar{Q}}=\mathbf{Q}\tan\delta_{Q}$, and $\mathbf{\bar{R}}=\mathbf{R}\tan\delta_{R}$, we obtain the equations
\begin{equation}\label{Eq:matrixStokesparameters}
\left( \matrix{
\bar{P}_{1} & \bar{P}_{2} & \bar{P}_{3} \cr \bar{Q}_{1} & \bar{Q}_{2} & \bar{Q}_{3} \cr \bar{R}_{1} & \bar{R}_{2} & \bar{R}_{3}
}\right) \left( \matrix{
A_{1} \cr A_{2} \cr A_{3} 
}\right) = \left( \matrix{
\tan (\Phi_{\mathrm{G}})_{1} \cr \tan (\Phi_{\mathrm{G}})_{2} \cr \tan (\Phi_{\mathrm{G}})_{3}
}\right).
\end{equation}
By solving the previous equations, we determine the Stokes parameters of the unknown polarisation state $\mathbf{A}$. The set of polarisation systems that we use to solve equation (\ref{Eq:matrixStokesparameters}) are
\begin{enumerate}
    \item $\mathbf{QWP}(0)$ with eigenvector $(1;0;0)$ and eigenvalue $\exp(i\pi/4)$.
    \item $\mathbf{QWP}(\pi/4)$ with eigenvector $(0;1;0)$ and eigenvalue $\exp(i\pi/4)$.
    \item $\mathbf{HWP}(\pi/8)\mathbf{HWP}(\pi/2)$ with eigenvector $(0;0;1)$ and eigenvalue $\exp(i\pi/4)$.
\end{enumerate}
Notice that all the systems have the same eigenvalues, hence $\delta_{P}=\delta_{Q}=\delta_{R}=\pi/4$. With $\tan(\pi/4)=1$, we find that
\begin{equation}\label{Eq:matrixStokessimple}
\left( \matrix{
A_{1} \cr A_{2} \cr A_{3} 
}\right) = \left( \matrix{
\tan (\Phi_{\mathrm{G}})_{1} \cr \tan (\Phi_{\mathrm{G}})_{2} \cr \tan (\Phi_{\mathrm{G}})_{3}
}\right).
\end{equation}

It is interesting to observe the similarities between the intensity-based Stokes parameters and the phase-based versions. The former relies on three intensity projections (assuming normalised Stokes parameters): one projection on the horizontal state, whose Stokes vector is $(1;0;0)$; another projection on the diagonal state with Stokes vector $(0;1;0)$; and the third projection on a circular polarisation state. Assuming right-handed circular polarisation state, the Stokes vector is $(0;0;1)$. Therefore, in comparison with the procedure listed above to measure the phase-based Stokes parameters, it is clear that we have found an alternative strategy based on measuring the geometric phase.

It is worth noticing that, with our current experimental procedure, it is impractical to determine the Stokes parameters using the above theory. Our experiment requires two orthogonal polarisation states. If the objective is to determine the Stokes parameters of an unknown polarisation state, it means that we need to find the orthogonal polarisation. Since, for any input state, there is no single element that generates the orthogonal polarisation, our current procedure is ineffective. Nevertheless, the above theory does not depend on our experimental method, and thus remains as an alternative formulation to determine the Stokes parameters. 


 
 \section{Fringe-contrast polarimetry}\label{Sec:fringe_contrast}
 
This method consists of using the visibility of the interference fringes to infer the geometric phase \cite{Garza-Soto2020}. In the previous method, we neglected the fringes' visibility. However, visibility plays an important role in the accuracy of the previous method. If the visibility is low, such that the oscillatory modulation of the intensity pattern disappears, the relative displacement between the fringe patterns is not measurable. For example, there are cases where the polarisation state exiting the polarising optical system is orthogonal to the input state. Consequently, the fringe pattern disappears, and hence, the previous method, which relies on the displacement of the interference fringes, is not suitable. These cases have been named ortho-transmission states \cite{Gutierrez-Vega2020-1}. 

\begin{figure}[ht!]
\centering\includegraphics[width=8cm]{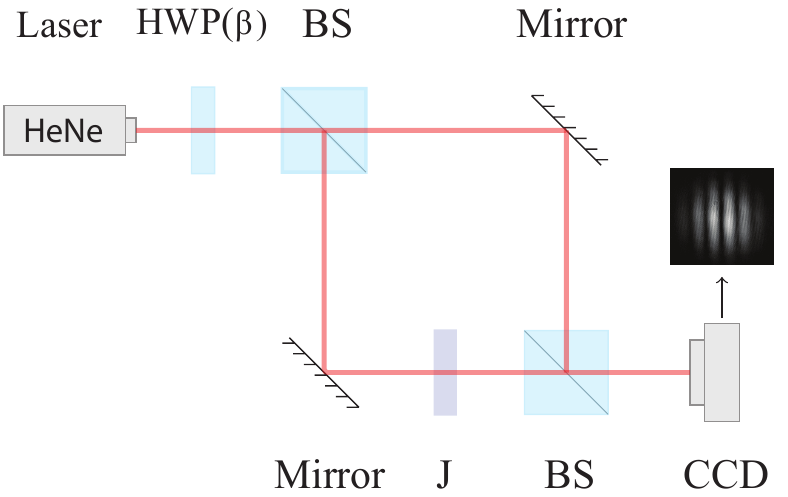}
\caption{Schematic of our experimental arrangement to perform fringe-contrast polarimetry. Laser source: linearly-polarised He-Ne laser. $\mathbf{HWP}(\beta)$: half-wave plate with fast axis oriented an angle $\beta$. It is used to controlled the input linear polarisation state to the Mach-Zehnder interferometer. \textbf{BS}: beam-splitter cubes. $\mathbf{J}$: polarising optical system. \textbf{CCD}: camera.}\label{Fig:experimentVis}
\end{figure}

Here, we use an MZ interferometer, as shown in figure \ref{Fig:experimentVis}. In comparison with the previous setup, this one uses cube BSs and a single interference pattern. Furthermore, we use linear polarization states as our input states, which are generated using a half-wave plate, $\mathbf{HWP}(\beta)$, before the interferometer. Since the laser is horizontally polarised, the output of $\mathbf{HWP}(\beta)$ is linearly polarised with inclination angle $\theta=2\beta$. The first BS creates two copies of the input polarization state $\mathbf{E}_{a}$. One copy travels through the upper arm of the interferometer and is used as the reference beam. The second copy goes to the lower arm and passes through the sample characterised by $J$. Let us consider $\mathbf{E}_{b}$ as the output Jones vector containing the information of the sample. In order to create the fringe interference pattern, the upper arm of the interferometer introduces a small tilt, so that the output reference beam can be written as $\mathbf{E}_{b}\exp (i\phi)$, where $\phi$ encompasses the transverse wavevector introduced by the tilt and the axis perpendicular to the oscillations of the fringes.

The interference fringes, $I(\phi)$,  resulting from the superposition of $\mathbf{E}_{a}$ and $\mathbf{E}_{b}\exp(i\phi)$ at the output of the interferometer is given by
\begin{equation}
    I(\phi)=|\mathbf{E}_{a}+\mathbf{E}_{b}e^{i\phi}|^2 = |\mathbf{E}_{a}|^2 + |\mathbf{E}_{b}|^2 + 2\mathrm{Re}(\mathbf{E}_{a} \cdot \mathbf{E}_{b}^{\ast} \, e^{-i\phi}). \label{Eq:inter_vis}  
\end{equation}
According to the standard definition of visibility, i.e., 
\begin{equation}
V=\frac{I_{\mathrm{max}} - I_{\mathrm{min}}}{I_{\mathrm{max}} + I_{\mathrm{min}}}, \label{Eq:visibility}
\end{equation}
the visibility is determined by finding the maximum ($I_{\mathrm{max}}$) and minimum ($I_{\mathrm{min}}$) values of $I(\phi)$. To do so, we substitute the dot product $\mathbf{E}_{a} \cdot \mathbf{E}_{b}^{\ast}$, in equation (\ref{Eq:inter_vis}), by 
\begin{equation}
    \mathbf{E}_{a} \cdot \mathbf{E}_{b}^{\ast} = \mu_1 + \mu_2 + (\mu_1 - \mu_2)\mathbf{Q}\cdot \mathbf{A}, 
\end{equation}
which comes from equation (\ref{Eq:totalphase}) [as a reminder, $\mathbf{A}$ is the Stokes vector of $\mathbf{E}_a$, and $\{ \mathbf{Q}, -\mathbf{Q}\}$, $\{\mu_1 , \mu_2\}$ are the Stokes eigenvectors and eigenvalues of $\mathbf{J}$, respectively]. In this manner, it is straightforward to show that 
\begin{equation}
    V=\sqrt{\cos^2 \delta  + (\mathbf{Q}\cdot \mathbf{A})^2 \sin^2 \delta}, \label{Eq:V1}
\end{equation}
where $\delta=\mathrm{arg}[\mu_1]$ (i.e., $\mu_1=\exp(i\delta)$ and $\mu_2=\mu_1^{\ast}$, as explained in section \ref{Sec:theory}). Moreover, we can rewrite equation (\ref{Eq:V1}) in terms of the geometric phase using equation (\ref{Eq:tanphiPB}), i.e.,
\begin{equation}
    V=\cos \delta \sqrt{1  + \tan ^2 \Phi_{\mathrm{G}}}. \label{Eq:V2}
\end{equation}
Equations (\ref{Eq:V1}) and (\ref{Eq:V2}) constitute the second important result of this work. They provide simple expressions connecting the visibility of the interference fringes, which is obtained from experimental measurements, with the geometric phase and the parameters of the Jones matrix characterising the sample. Therefore, (\ref{Eq:V1}) and (\ref{Eq:V2}) are the main equations behind our fringe-contrast polarimetry technique.

We remark that we are not implying that the geometric phase $\Phi_{\mathrm{G}}$ changes the visibility of the fringes. $\Phi_{\mathrm{G}}$ is a global phase, and thus it only produces a displacement of the fringe pattern according to the explanation of section \ref{Sec:fringe_shifting}. The visibility changes because $\mathbf{E}_b$ exits the sample with a different polarization state from $\mathbf{E}_a$ (except for the cases when $\mathbf{E}_a$ is an eigenvector). However, since $\mathbf{E}_{b}$ is related to $\Phi_{\mathrm{G}}$ by equation (\ref{Eq:totalphase1}), we can connect the visibility and the geometric phase, as given by equation (\ref{Eq:V2}).

Finally, for completeness, an equivalent formulation of equation (\ref{Eq:V2}) in terms of the input Jones vector $\mathbf{E}_a$ is given by
\begin{equation}
    V=\cos(\Phi_{\mathrm{G}}-\delta) (\mathbf{E}_1\cdot\mathbf{E}_a)^2 + \cos(\Phi_{\mathrm{G}}+\delta) (\mathbf{E}_2\cdot\mathbf{E}_a)^2,
\end{equation}
where $\{ \mathbf{E}_1 , \mathbf{E}_2 \}$ are the Jones eigenvectors of $\mathbf{J}$.

 
\subsection{Proof-of-concept experiment} 
Our proof-of-principle experiment follows the schematic presented in figure \ref{Fig:experimentVis}. The half-wave plate allows us to generate any linear state of polarization. It is controlled using a motorized rotation stage (Thorlabs KPRM1E). The beam passes through an MZ interferometer which is aligned as explained in the previous section to generate a fringe pattern at the output. Notice that in this method, since we are only concern about the contrast of the fringes, we do not need to compensate for the dynamic phase. The resulting interference fringes are recorded using a CCD camera. Similar to the previous method, the images pass through a numerical low-pass filter to eliminate noise introduced by the infrared filter of the camera, multiple unwanted reflections, dust and imperfections on the optical elements. 

To determine the visibility of the interference fringes, we find, numerically, the maximum and minimum intensities and apply equation (\ref{Eq:visibility}). However, we note that the interference fringes are apodised by the envelope of the laser beam's transverse profile. Therefore, the intensity of the fringes decreases away from the optical axis of the beam. Thus, it is important to have several thin fringes (by tilting the reference mirror) to mitigate the complications introduced by the envelope. Taking that into account, in our numerical method, we locate the local maximum and minimum intensities closer to the centre of the beam.   

The Jones matrix that we test consists of a $\mathbf{QWP}(45^{\circ})$ sandwiched between two $\mathbf{HWP}(0^{\circ})$ whose fast axes are both oriented parallel to the horizontal polarization. This configuration has diagonal and antidiagonal eigenpolarisations [i.e., $\mathbf{Q}_{1,2}=(0;\pm 1;0)$], and introduces a retardance equal to $\pi$ [see equation (\ref{Eq:retardance})]. Using equation (\ref{Eq:V1}) and linear polarisation states to probe the system [i.e., $\mathbf{A}=(\cos 2\theta;\sin 2\theta;0)$], the expected visibility becomes
\begin{equation}
    V=|\sin 2\theta|. \label{Eq:Vtheoryexp}
\end{equation}
For the derivation of the above result, notice that $\delta=\pi/2$ [cf. equation (\ref{Eq:eigenvaluesQHQ})] and $\theta = 2\beta$ is the inclination angle of the input polarisation ($\beta$ is the angle of the $\mathbf{HWP}$ that we rotate before the MZ interferometer).

\begin{figure}[ht!]
\centering\includegraphics[width=8cm]{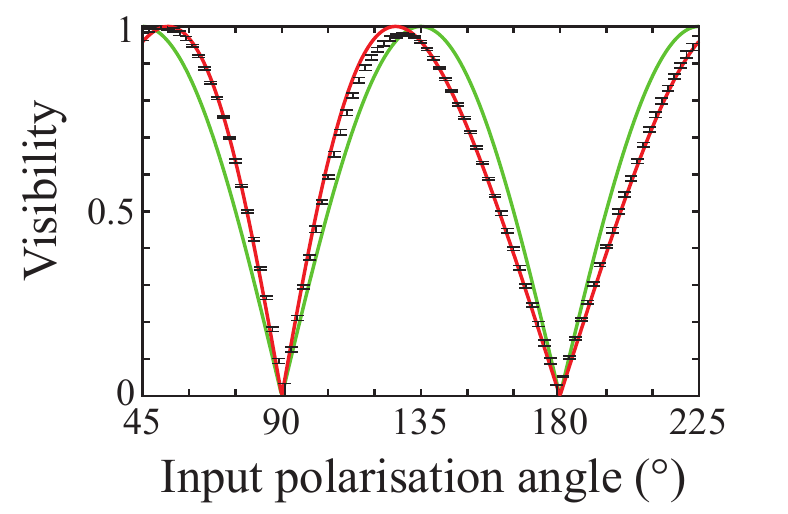}
\caption{Results of our fringe-contrast polarimetry experiment. Here, we show the visibility of the interference fringes using a homogeneous system made of a half-wave plate sandwiched between two quarter-wave plates. We test the experiment with linear states from the diagonal polarisation ($\theta = 45^{\circ}$) to a linear state with inclination angle $\theta = 225^{\circ}$. The error bars show the experimental measurements. The green line is the theoretical result [cf. equation (\ref{Eq:Vtheoryexp})] assuming an experiment with ideal 50:50 beam splitters. The red line is the numerical result considering asymmetric beam splitters. See the main text for further details. }\label{Fig:visresults}
\end{figure}

Figure \ref{Fig:visresults} shows the results of our experiment. We initialized the state generator $\mathbf{HWP}$ at $\beta=\pi/8$, and thus the input polarisation is diagonal and corresponds to one of the eigenpolarisations. Hence, we expect maximum visibility at this initial position. The error bars in the plot correspond to the experimental measurements. We used angular steps of $2^\circ$, and repeated the experiment ten times. The green line is the theoretical result given by equation (\ref{Eq:Vtheoryexp}). As noted in the results, there is a mismatch between the theoretical curve and the measurements. We explain and demonstrate below the reason for the discrepancy. Nevertheless, the ideal theory and the experiment agree in the principal features, i.e., we observed maximum visibility near the eigenpolarisations of the system and minimum visibility in the intermediate states.

The discrepancy between the experiment and the theory is consequence of our non-ideal BSs. Let us consider asymmetric BSs where the reflection ($r_{\mathbf{s},\mathbf{p}}$) and transmission ($t_{\mathbf{s},\mathbf{p}}$) coefficients vary between the $\mathbf{s}$- and $\mathbf{p}$-polarised modes. Here, the $\mathbf{s}$ and $\mathbf{p}$ modes match our vertical and horizontal polarisations, respectively. Therefore, the reflected and transmitted beams from a BS are written as
\begin{eqnarray}
\mathbf{E}_r &=& r_\mathbf{s} \mathbf{E}_\mathbf{s} + r_\mathbf{p} \mathbf{E}_\mathbf{p}, \\
\mathbf{E}_t &=& t_\mathbf{s} \mathbf{E}_\mathbf{s} + t_\mathbf{p} \mathbf{E}_\mathbf{p},
\end{eqnarray}
where $\mathbf{E}_\mathbf{s}$ and $\mathbf{E}_\mathbf{p}$ are the $\mathbf{s}$ and $\mathbf{p}$ components of the input polarisation. By taking the asymmetric BSs into account, we obtain, using Jones calculus, the red solid line shown in figure \ref{Fig:visresults}. The values that we consider for the reflectance ($R_{s,p}=|r_{s,p}|^2$) and transmittance ($T_{s,p}=|t_{s,p}|^2$) of both BSs are $\{ R_{s},R_{p}\}=\{0.68,0.38\}$ and $\{T_{s},T_{p}\}=\{0.27,0.59\}$, respectively. Notice that $R_{s,p}+T_{s,p}<1$, which is attributed to absorption losses. The agreement with the experiment makes clear that the asymmetric BSs are responsible of the mismatch between our ideal theory -- which considers 50:50 BSs -- and the experiment. One might be tempted to incorporate the effect of the asymmetric BS into our formulation to determine an analytical expression for the visibility similar to equation (\ref{Eq:Vtheoryexp}). However, our formulation is strictly considering homogeneous Jones matrices. If we introduce the asymmetric BS as part of the system's Jones matrix, the resulting matrix becomes inhomogeneous, and hence, our formulation is not applicable.

\section{Discussion and conclusions}\label{Sec:conclusions}
 
We have shown two polarimetry techniques based on the geometric phase. These methods, to the best of our knowledge, are novel and can be classified in a new category named: geometric-phase polarimetry (GPP). The applications of this technique are twofold: (i) it can be used to obtain the Jones matrix properties of a polarising optical system and (ii) it can be used to measure the Stokes parameters of a light beam.

The first GPP technique that we demonstrated uses the relative displacement between two interference fringe patterns obtained using an MZ interferometer. One of the interference fringes corresponds to a generic elliptical polarisation state entering the interferometer, while the second interferogram corresponds to the orthogonal polarisation state. We tested our technique by measuring the retardance introduced by a homogeneous Jones matrix. The experimental and theoretical results are in agreement with our formulation.

The second GPP technique relates to the visibility of the interference fringes with the geometric phase. This method offers a significant advantage in comparison with many experiments that measure the geometric phase through interferometry: it does not require compensating the dynamic phase. We have experimentally demonstrated this technique by measuring the visibility in terms of the input polarisation to the system. Although the results of the experiment appear to be in disagreement with the ideal theory, we showed that this is due to the asymmetric beam splitters used in the interferometer. Therefore, this method can be improved by using symmetric beamsplitters. 

Of course, it is questionable if these methods, which require interferometric measurements, are feasible in a practical setting. However, the theory described in section \ref{Sec:theory} is independent of the experiment used to measure the geometric phase and is valid as long as we can accurately measure $\Phi_{\mathrm{G}}$. Therefore, other techniques to measure the geometric phase can be used; for example, Malhotra \emph{et al.} \cite{Malhotra2018} proposes a method that does not require interferometry. 

Furthermore, GPP can be applied to other physical systems, equally described by the SU(2) group. For example, time-of-flight spectrometry is used to characterised particle beams (e.g. ion or electron beams). This technique measures the time taken by the particle to travel a distance through a medium. If the medium is anisotropic, the particle beam acquires a geometric phase that increases the time of flight. Therefore, we can use the time-of-flight information with GPP to characterise polarised particle beams \cite{KARIMI201422}.

Finally, inhomogeneous Jones matrices are more involved since the geometric phase between orthogonal polarisation states is no longer opposite in sign but completely different, and therefore, our experiment cannot be applied. Future experimental work is devoted to the characterisation of inhomogeneous Jones matrices \cite{Gutierrez-Vega2020-3}.

\section*{Acknowledgements}

Dorilian Lopez-Mago acknowledges support from Consejo Nacional de Ciencia y Tecnolog\'{i}a (CONACYT) through the grants: 257517, 280181 and 293471.

\section*{References}

\bibliographystyle{iopart-num}
\bibliography{bibliography}

\end{document}